\pdfoutput=1
\let\ifpdf\relax

\documentclass{PoS}
\let\ifpdf\relax

\newcommand{\bd}{\begin{displaymath}}
\newcommand{\ed}{\end{displaymath}}
\newcommand{\be}{\begin{equation}}
\newcommand{\ee}{\end{equation}}

\usepackage{capt-of}

\title{Nonperturbative QCD corrections to electroweak observables}

\ShortTitle{Nonperturbative QCD corrections to electroweak observables}

\author{\speaker{Dru B.\ Renner}\\
Thomas Jefferson National Accelerator Facility (JLab)\\
E-mail:\ \email{dru@jlab.org}}

\author{Xu Feng\\
High Energy Accelerator Research Organization (KEK)}

\author{Karl Jansen\\
Deutsches Elektronen-Synchrotron (DESY)}

\author{Marcus Petschlies\\
The Cyprus Institute}

\abstract{Nonperturbative QCD corrections are important to many
  low-energy electroweak observables, for example the muon magnetic
  moment. However, hadronic corrections also play a significant role
  at much higher energies due to their impact on the running of
  standard model parameters, such as the electromagnetic
  coupling. Currently, these hadronic contributions are accounted for
  by a combination of experimental measurements, effective field
  theory techniques and phenomenological modeling but ideally should
  be calculated from first principles.  Recent developments indicate
  that many of the most important hadronic corrections may be feasibly
  calculated using lattice QCD methods.  To illustrate this, we will
  examine the lattice computation of the leading-order QCD corrections
  to the muon magnetic moment, paying particular attention to a
  recently developed method but also reviewing the results from other
  calculations.  We will then continue with several examples that
  demonstrate the potential impact of the new approach:\ the
  leading-order corrections to the electron and tau magnetic moments,
  the running of the electromagnetic coupling, and a class of the
  next-to-leading-order corrections for the muon magnetic moment.
  Along the way, we will mention applications to the Adler function,
  which can be used to determine the strong coupling constant, and QCD
  corrections to muonic-hydrogen.}

\FullConference{XXIX International Symposium on Lattice Field Theory\\
July 10 - 16, 2011\\
Squaw Valley, Lake Tahoe, California}

\begin{document}

\section{Introduction}

Many precision experiments are increasingly becoming sensitive to
nonperturbative QCD corrections.  For example, the measurement of the
magnetic moment of the muon currently shows a discrepancy with the
standard model of over three standard deviations.  Its theoretical
uncertainty is dominated by hadronic effects, making the muon
$g\,$-$\,2$ a prominent example of the importance of a fully
nonperturbative determination of hadronic corrections.  This is just
one example.  The significance of QCD corrections to otherwise
precision observables will increase with the experimental programs
envisioned for the future.  In many cases, the discovery of physics
beyond the standard model may depend on accurate control of these
hadronic effects.

In these proceedings, we will discuss several opportunities for
lattice QCD calculations of hadronic corrections to important
measurements that may in fact be more feasible than previously
thought.  The QCD corrections to the muon $g\,$-$\,2$ will serve as a
concrete example.  This will allow us to identify an issue that makes
the calculation of these quantities more difficult and then describe a
modified method that was introduced to alleviate this
problem~\cite{Feng:2011zk}.  Additionally, there is a growing lattice
effort on precisely this quantity and we will review the latest
results.

After having laid the groundwork with the muon $g\,$-$\,2$, we will
continue with several additional examples that illustrate the
potential impact of the modified technique.  We will examine the
leading-order hadronic corrections for the electron and tau magnetic
moments and the leading QCD contributions to the running of the QED
coupling.  Additionally, we will note applications to the Adler
function, the determination of the strong coupling constant and the
QCD corrections to the energy levels of muonic-hydrogen.  We will then
close with a calculation of the next-to-leading-order
vacuum-polarization corrections to the muon $g\,$-$\,2$.

\section{Leading-order QCD correction to the muon magnetic moment}

The BNL measurement of the anomalous magnetic moment of the muon
$a_\mu=(g_\mu-2)/2$~\cite{Bennett:2004pv} and the standard model
estimate thereof~\cite{Jegerlehner:2009ry} differ by more than three
standard deviations.  This discrepancy may indicate physics beyond the
standard model, but making such a statement definitively requires a
thorough understanding of all sources of uncertainty and ideally a
significantly larger discrepancy.  The experimental community is
pursuing two future muon $g\,$-$\,2$ experiments at
Fermilab~\cite{Roberts:2010cj} and J-PARC~\cite{Toyoda:2011qs}, aiming
to improve the experimental precision on $a_\mu$ from $6.3\,\cdot
10^{-10}$ to $(1-2)\,\cdot 10^{-10}$.  Since $a_\mu\approx 1.2\cdot
10^{-3}$, the new experiments will reduce the relative precision from
$0.5\cdot 10^{-6}$ to $0.9\cdot 10^{-7}$.  At this precision, the
comparison between theory and experiment would be dominated by the
standard model uncertainties alone, hence improvement from the theory
side is highly desirable.

\begin{figure}
\begin{minipage}{200pt}\vspace{12pt}\vspace{-4pt}
\begin{center}
{\small
\begin{tabular}{|l|r|}\hline
Contribution & Error $[10^{-10}]$ \\\hline
QCD-LO & 5.3 \\
QCD-NLO & 3.9 \\
QED/EW & 0.2 \\\hline
Total & 6.6 \\\hline
\end{tabular}\vspace{0pt}
}
\end{center}
\captionof{table}{Standard model uncertainties in
  $a_\mu$~\cite{Jegerlehner:2009ry}.  The contributions QCD-LO and
  QCD-NLO refer to $a_\mu^{(2)}$ and $a_\mu^{(3)}$ in
  equation~\protect\ref{amuexp}.  All remaining contributions are
  collected together and labeled QED/EW.}
\label{error-budget}
\end{minipage}
\hspace{26pt}
\begin{minipage}{200pt}
\begin{center}
\includegraphics[width=90pt]{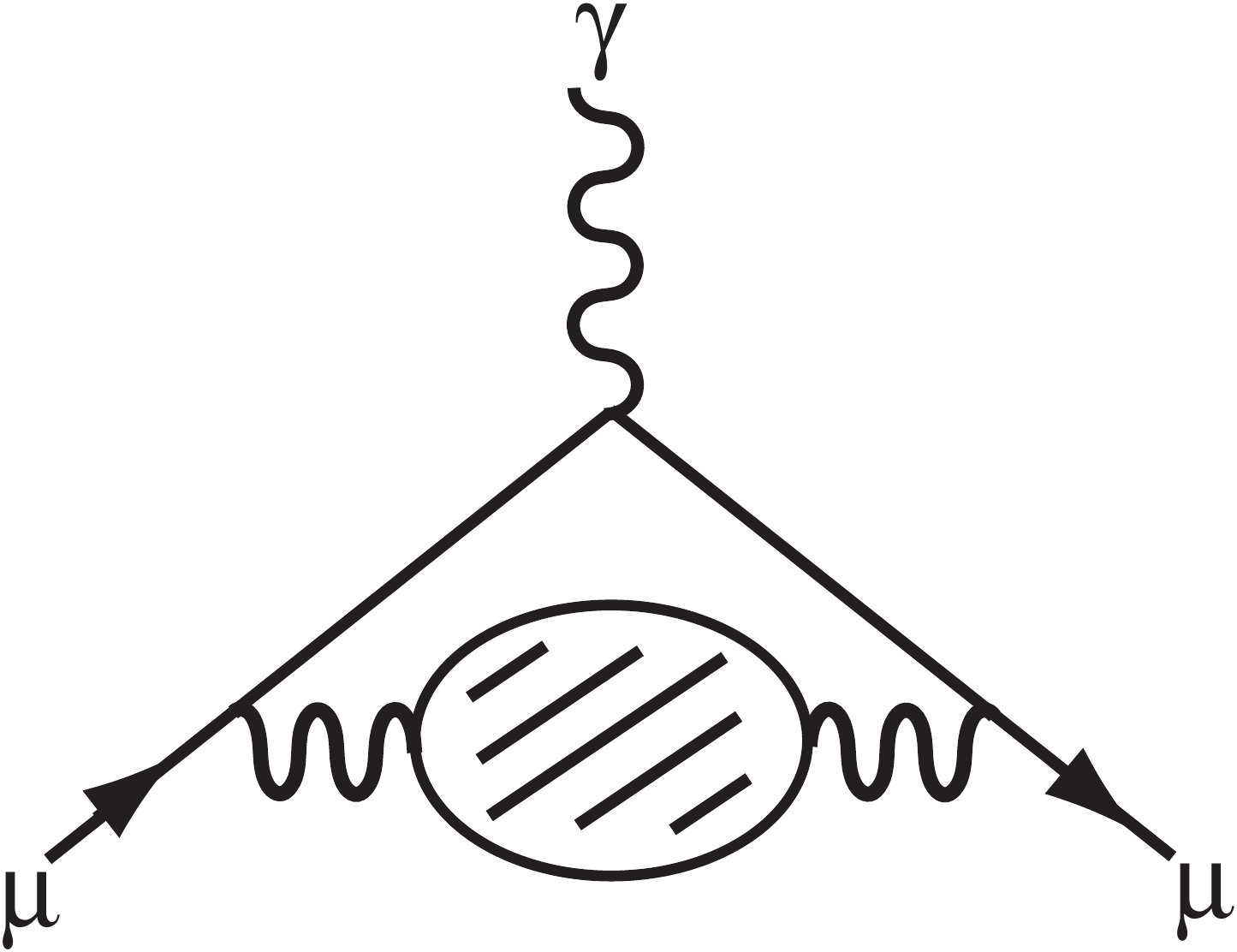}
\end{center}
\caption{Leading-order QCD contribution $a_\mu^{(2)}$.  The QCD
  contribution to $a_\mu^{(2)}$, denoted by the shaded region, can be
  related to $R(s)$, given in equation~\protect\ref{al2-disp},
  determined experimentally or calculated using lattice QCD.}
\label{lo-diag}
\end{minipage}
\end{figure}

The value of $a_\mu$ receives contributions from all parts of the
standard model, each contributing to the theoretical uncertainty as
shown in table~\ref{error-budget}.
Quite clearly, the standard model uncertainty is overwhelmingly
dominated by hadronic physics.  The total QCD contribution
$a_\mu^\mathrm{QCD}$ can be organized as an expansion in the QED
coupling $\alpha$ as follows
\be
\label{amuexp}
a_\mu^\mathrm{QCD}
= \alpha^2 A_\mu^{(2)} + \alpha^3 A_\mu^{(3)} + {\cal O}(\alpha^4)
= a_\mu^{(2)} + a_\mu^{(3)} + {\cal O}(\alpha^4)\,,
\ee
where $a_\mu^{(n)}= \alpha^n A_\mu^{(n)}$.  The expansion in $\alpha$
is perturbative but the $A_\mu^{(n)}$ must be calculated
nonperturbatively.  The largest source of uncertainty is due to
$a_\mu^{(2)}$, which we discuss next.

\subsection{Experimental determination of $\bf{a_\mu^{(2)}}$}

The leading-order correction $a_\mu^{(2)}$, shown in
figure~\ref{lo-diag}, can be determined experimentally.  It can be
written as an integral of $R(s)$ and a known function $K(s/m_\mu^2)$
as~\cite{Jegerlehner:2009ry}
\be
\label{al2-disp}
a_\mu^{(2)}
= \alpha^2 \int_{4m_\pi^2}^\infty \frac{ds}{s} K(s/m_\mu^2) R(s)
~~~~\mathrm{with}~~~~
R(s)
= \frac{\sigma(\gamma^\ast\rightarrow\mathrm{hadrons})}{\sigma(\gamma^\ast\rightarrow e^+e^-)}\,.
\ee
In practice, measurements of
$\sigma(\gamma^\ast\rightarrow\mathrm{hadrons})$ from many different
experiments are combined to form the integral above.  The most recent
compilation of measurements results in $a_\mu^{(2)} =
6.923\,(42)\cdot10^{-8}$~\cite{Davier:2010nc}, which is an improvement
on the error given in table~\ref{error-budget}. This approach has been
and will continue to be for some time the most accurate means of
providing the QCD input needed to form the standard model prediction
for $a_\mu$.  This value is significantly more precise than current
lattice results, but we should bear in mind that this result requires
a substantial experimental effort to determine a quantity that should
in principle be predicted from the theory itself.  Reaching and even
exceeding the precision of the experimental determination of
$a_\mu^{(2)}$ should be part of the long-term efforts of the lattice
community.

\subsubsection{Estimates of the flavor dependence of $\bf{a_\mu^{(2)}}$}

As we will see shortly, lattice calculations must include the four
lightest quarks to reach the precision on $a_\mu^{(2)}$ needed for the
future muon $g\,$-$\,2$ experiments.  However, the number of quark
flavors used in current lattice computations varies from $n_f= 2$ to
$3$ and just recently $4$ flavors.  Thus it is useful to have some
guidance on the $n_f$ dependence of $a_\mu^{(2)}$.  Unfortunately,
there is no unique way to do this for $a_\mu^{(2)}$, which is a
problem for many other observables as well.  The effects of decoupling
a quark flavor depend on the renormalization conditions used for the
remaining degrees of freedom.  This ambiguity exists in perturbation
theory and would also apply to any analysis of the experimental
results.  The advantage of lattice calculations is that they can
provide a well-defined way to prescribe such a definition.

Keeping in mind these limitations, we proceed with a simple means of
estimating the flavor contributions from the experimental
measurements.  We start with the experimentally determined
$R(s)$~\cite{fred} and rescale by the electric charges $Q_f$ of the
relevant quark flavors $f$ as follows
\be
\label{amunf}
R_{n_f}(s)
\equiv R(s) \frac{ \sum_{f}^{n_f} Q_f^2 }{ \sum_f^n Q_f^2}~~~~\mathrm{for}~~~~ 4m_n^2\le s \le 4 m_{n+1}^2
~~~~\mathrm{with}~~~~ 
a_{\mu,n_f}^{(2)}
\equiv \alpha^2 \int_{4m_\pi^2}^\infty \frac{ds}{s} K(s/m_\mu^2) R_{n_f}(s)\,.
\ee
The $\sum_f^{n}$ accounts for the $n$ degrees of freedom present in
the experimental measurements between the quark thresholds $4 m_n^2$
and $4 m_{n+1}^2$ and the $\sum_f^{n_f}$ restores the desired $n_f$
flavors.  The leading-order perturbative contribution to $R(s)$ scales
this way, so this prescription is valid up to perturbative corrections
and away from the resonance regions.  Near resonances, this amounts to
a quark-hadron duality argument.  For example, the strange quark
content of the prominent $\phi$ meson would not follow such a scaling
but the integrand $K(s/m_\mu^2)$ is relatively smooth and effectively
averages $R(s)$ across a window in $s$ over which this scaling is
expected to be more effective.  In practice, the part of this
prescription that has the most impact on the resulting estimates of
$a_{\mu,n_f}^{(2)}$ is the use of quark masses rather than a resonance
mass to define the thresholds.  And in fact, only the strange quark
threshold is particularly sensitive to this choice.

\begin{figure}
\begin{minipage}{210pt}
\includegraphics[width=210pt]{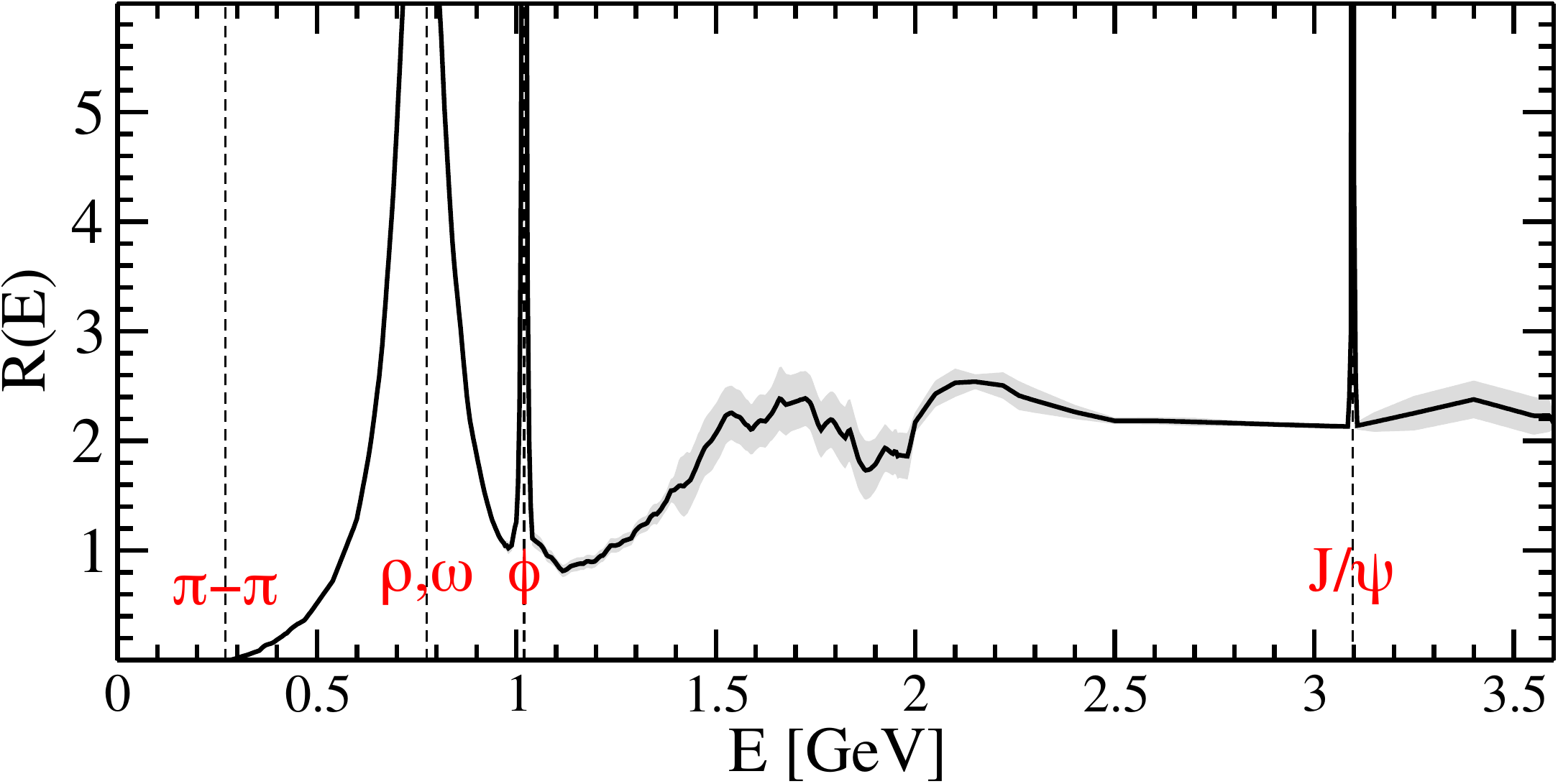}
\caption{Measured $R(E)$.  The $R(s)$ with $\sqrt{s}=E$ from
  F.\ Jegerlehner~\cite{fred} is shown.}
\label{rs}
\end{minipage}
\hspace{13pt}
\begin{minipage}{203pt}
\begin{center}
{\small
\begin{tabular}{|c|c|c|} \hline
$n_f$ & $a_{\mu,n_f}^{(2)}$ [Exp] & $a_{\mu,n_f}^{(2)}$ [Lat] \\\hline
$5$ & $6.93\,(06)$ & - \\\hline
$4$ & $6.93\,(06)$ & underway~\cite{grit} \\\hline
$3$ & $6.81\,(05)$ & $6.41\,(46)$~\cite{Boyle:2011hu}, $6.18\,(64)$~\cite{DellaMorte:2011aa} \\\hline
$2$ & $5.67\,(05)$ & $5.72\,(16)$~\cite{Feng:2011zk}, $5.46\,(66)$~\cite{DellaMorte:2011aa} \\\hline
\end{tabular}
}
\end{center}
\captionof{table}{Estimated $n_f$ dependence of $a_\mu^{(2)}$ (given
  in units of $10^{-8}$) from experimental measurements using
  equation~\protect\ref{amunf} ("[Exp]") and current lattice
  calculations ("[Lat]").}
\label{nf-dep}
\end{minipage}
\end{figure}

Using this simple prescription, the $n_f$ dependence of $a_\mu^{(2)}$
can be estimated.  To do this, we have used the $R(s)$ compiled by
F.\ Jegerlehner~\cite{fred}.  This is shown in figure~\ref{rs}.
The results of this procedure and a comparison with the current
lattice calculations are given in table~\ref{nf-dep}.  This suggests
that the charm quark correction is roughly $1.2\cdot 10^{-9}$.  The
current experimental precision is $6.3\cdot 10^{-10}$, which makes it
clear that charm quark contributions are already necessary to reach
the precision of the BNL measurement, let alone the precision for the
future $g\,$-$\,2$ experiments.

Alternatively, one could attempt to assign flavor weights to each of
the final states in the total cross section
$\sigma(\gamma^\ast\rightarrow\mathrm{hadrons})$ and form
$a_{\mu,n_f}^{(2)}$.  This approach is an alternative prescription to
the one given above and, in fact, differs from it~\cite{fred2,davier},
illustrating the ambiguity in extracting the $n_f=2$ piece.  It turns
out to lead to a larger $n_f=2$ contribution than given in recent
$n_f=2$ lattice calculations, but we want to emphasize that any such
ambiguities are systematically eliminated as the lattice calculations
account for all the relevant quark flavors, which appears to be
$n_f=4$.

\subsection{Lattice calculation of $\bf{a_\mu^{(2)}}$}

The standard method to calculate $a_\mu^{(2)}$ using lattice QCD was
given by Blum in~\cite{Blum:2002ii}.  It requires calculating the
vacuum-polarization function $\Pi(Q^2)$ and evaluating the integral
\be
\label{al2-lat}
a_\mu^{(2)} =
\alpha^2 \int_0^\infty \frac{dQ^2}{Q^2} \omega(Q^2/m_\mu^2) \Pi_R(Q^2)\,.
\ee
The weight function $\omega$ accounts for the perturbative portion of
the diagram in figure~\ref{lo-diag} and is known.  $\Pi_R(Q^2)$ is the
once-subtracted vacuum-polarization function, $\Pi_R(Q^2) \equiv
\Pi(Q^2) - \Pi(0)$, where $\Pi$ is given by
\bd
(Q_\mu Q_\nu - Q^2 \delta_{\mu\nu}) \Pi(Q^2) \equiv \int d^4X e^{iQ\cdot(X-Y)} \langle J_\mu(X) J_\nu(Y) \rangle\,,
\ed
and is directly calculable in Euclidean space.  This formulation can
be related to the approach used to experimentally determine
$a_\mu^{(2)}$ by noting that $R(s)$ is proportional to $\mathrm{Im}\,
\Pi(-s+i\epsilon)$, which is non-zero only on the branch cut starting
at $s=4m_\pi^2$.  A standard dispersion analysis then allows one to
relate equations~\ref{al2-disp} and \ref{al2-lat}.

\subsubsection{Role of external leptonic scale in $\bf{a_\mu^{(2)}}$}

Naively, the lattice calculation of $a_\mu^{(2)}$ should be relatively
easy.  The only non-trivial part of the computation is the calculation
of the Euclidean two-point correlation function $\langle J_\mu(X)
J_\nu(Y) \rangle$, which can be accurately determined.  Furthermore,
the quantity $a_\mu^{(2)}$ is dimensionless, so it seems reasonable to
expect that it may exhibit a relatively mild dependence on the scales
in the problem, particularly the quark masses and lattice spacing.
\begin{figure}
\begin{minipage}{191pt}
\includegraphics[width=191pt]{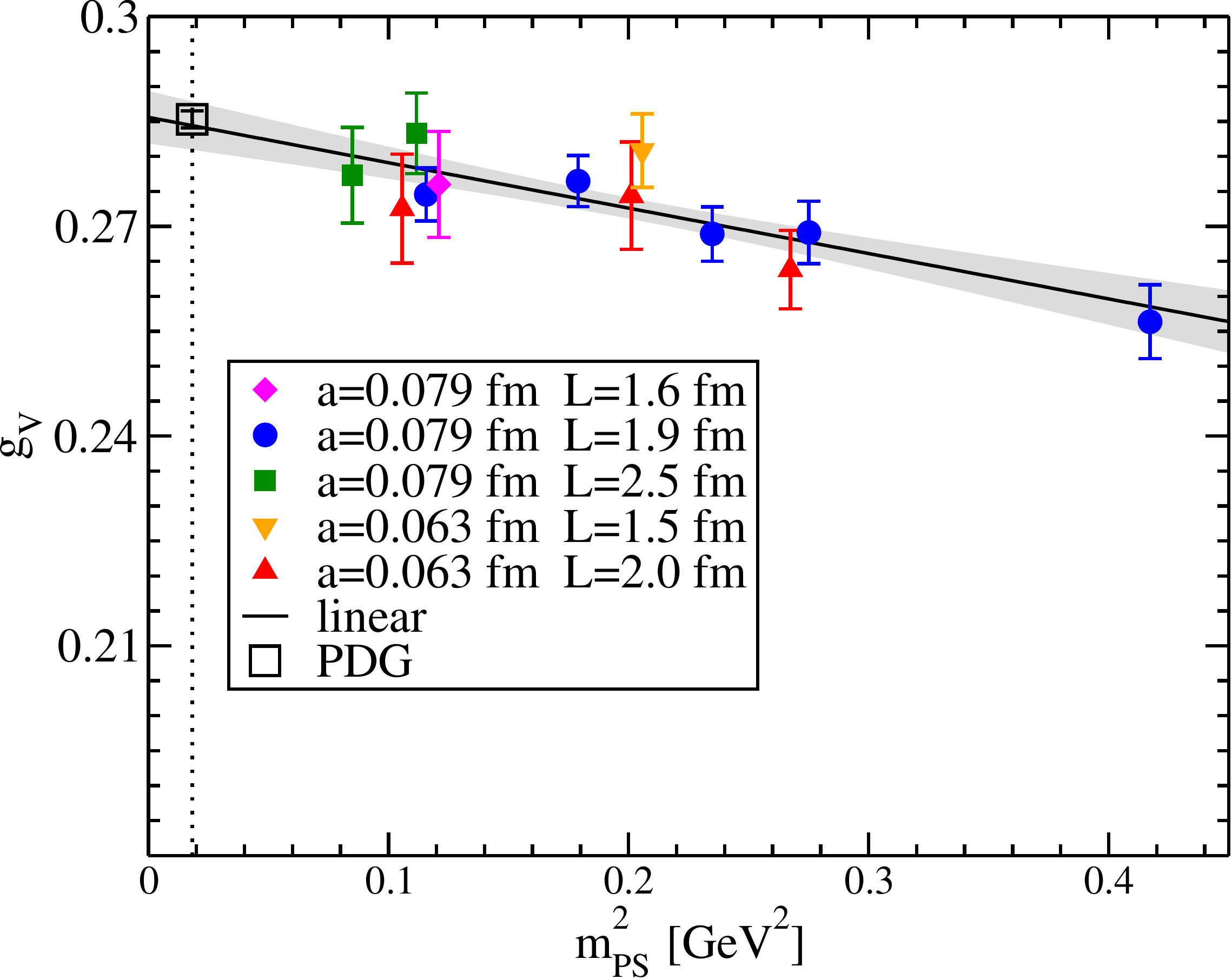}\vspace{-3pt}
\caption{Vector-meson coupling $g_V$.}
\label{gv}
\end{minipage}
\hspace{44pt}
\begin{minipage}{191pt}\vspace{5pt}
\includegraphics[width=191pt]{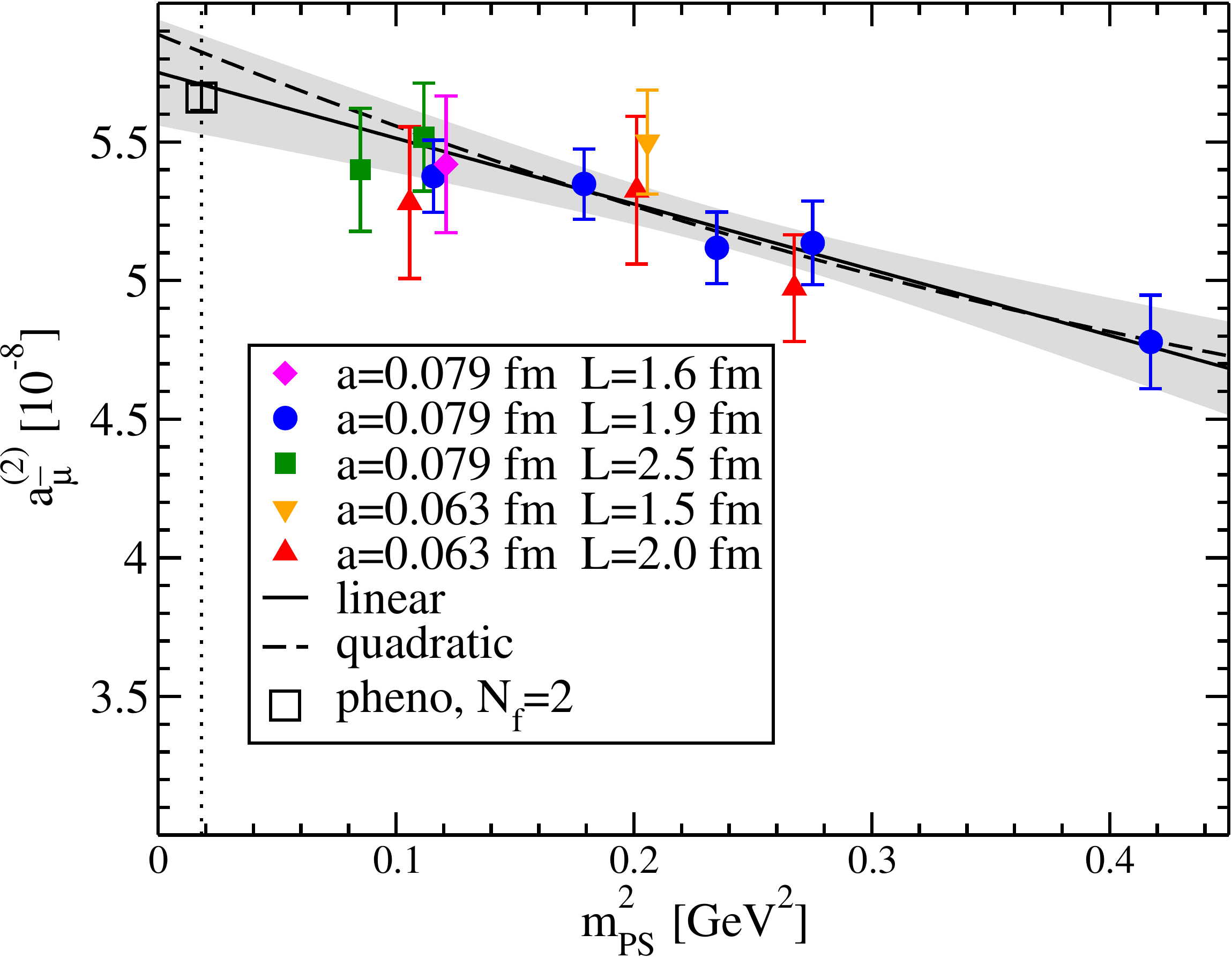}
\caption{Modified method $a_{\overline{\mu}}^{(2)}$.}
\label{amubar}
\end{minipage}
\end{figure}
(Figure~\ref{gv} shows an example of a typical dimensionless
quantity.)  However, $a_\mu^{(2)}$ is dimensionless only at the
expense of introducing an external scale, the muon mass $m_\mu$, and
this has several consequences.

First, $m_\mu$ introduces a dependence on the lattice spacing $a$ in
an otherwise dimensionless observable.  We can see this by writing the
integration variable for $a_\mu^{(2)}$ in lattice units
\bd
a_\mu^{(2)} =
\alpha^2 \int_0^\infty \frac{d\hat{Q}^2}{\hat{Q}^2} \omega(\hat{Q}^2/(am_\mu)^2) \Pi_\mathrm{lat}(\hat{Q}^2)
\ed
where $\hat{Q} \equiv a Q$ is the momentum variable in lattice units
and $\Pi_\mathrm{lat}(\hat{Q}^2)\equiv \Pi_R(\hat{Q}^2/a^2)$ is
directly calculated in lattice units.  Thus the lattice spacing $a$ is
needed in physical units to form $a m_\mu$.  This suggests that
$a_\mu^{(2)}$ may behave more like a dimensionful quantity.  As a
second consequence, the introduction of $m_\mu$ also allows for a
stronger quark mass dependence than might otherwise be expected.  The
dominant contribution of the lightest vector-meson with mass $m_V$ and
electromagnetic coupling $g_V$ is proportional to $g_V^2
m_\mu^2/m_V^2$.  This is a model-dependent statement, but it is
suggestive that $a_\mu^{(2)}$ may in fact behave more like a mass
dimension -$2$ observable.  (A plot of $m_V$ corresponding to the
$g_V$ in figure~\ref{gv} is given in~\cite{Feng:2011zk}.)

These two observations can be made precise by introducing an effective
dimension
\bd
d_\mathrm{eff}[X] \equiv
-\frac{a}{X} \left. \frac{\partial X}{\partial a}\right|_{g}\,.
\ed
This quantity is defined so that it is sensitive to only QCD scales
rather than the overall dimension.  To accomplish this, we write the
observable X as a function of both the lattice spacing $a$ and the
coupling $g$ separately $X=X(a,g)$.  Of course, $a=a(g)$ is eventually
chosen to be some function of the coupling, but treating $a$ and $g$
separately allows us to isolate the impact of the scale setting on the
quantity $X$.  Furthermore, $d_\mathrm{eff}$ is defined so that it
reproduces the usual definition of dimension for a simple QCD
observable but it differs for composite observables.

Several examples may help illustrate $d_\mathrm{eff}$.  First consider
some observable $M$ that is a standard QCD quantity of mass dimension
$n$.  This quantity would satisfy $M(a,g) = a^{-n} \hat{M}(g)$, where
$\hat{M}(g)$ is what is calculated on the lattice and the factor of
$a^{-n}$ is eventually put in by hand.  Then for such an $M$ we have
\bd
d_\mathrm{eff}[M] = 
- \frac{a}{a^{-n} \hat{M}(g)} \frac{\partial}{\partial a} \left(  a^{-n} \hat{M}(g) \right) = n\,.
\ed
Thus the dimension of quantities that we normally calculate is
unaltered.  However, for a composite object the answer can differ.
For example,
\bd
d_\mathrm{eff}\left[ g_V^2 m_\mu^2 / m_V^2 \right] = d_\mathrm{eff}\left[ \hat{g}_V^2 a^2 m_\mu^2 / \hat{m}_V^2 \right] = -2\,
\ed
where $g_V^2 m_\mu^2 / m_V^2$ is the leading piece of the vector-meson
contribution to $a_\mu$ mentioned earlier.  Thus $d_\mathrm{eff}$
captures the dimensionality of the QCD scales in this rather simple
composite expression.

We can now apply the definition of $d_\mathrm{eff}$ to $a_\mu^{(2)}$
and we find
\bd
d_\mathrm{eff}[a_\mu^{(2)}] =
-2 \left( \int_0^\infty \frac{dQ^2}{Q^2} \omega(Q^2/m_\mu^2) Q^2 \frac{d\Pi_R}{dQ^2} \right)
\left( \int_0^\infty \frac{dQ^2}{Q^2} \omega(Q^2/m_\mu^2) \Pi_R(Q^2) \right)^{-1}\,.
\ed
This quantity has a continuum limit.  It is rather easy to show that
$d_\mathrm{eff}[a_\mu^{(2)}]<0$, making it clear that $a_\mu^{(2)}$
acts like an observable with a negative mass dimension.  Additionally,
for $m_\mu\rightarrow 0$, we have $d_\mathrm{eff} \rightarrow -2$, and
for $m_\mu\rightarrow \infty$, we can show that $d_\mathrm{eff}
\rightarrow 0$.  For an intermediate mass, this quantity must be
calculated nonperturbatively.  For the muon, we find
\bd
d_\mathrm{eff}[a_\mu^{(2)}] = -1.887\,(5)\,
\ed
which clearly indicates that $a_\mu^{(2)}$ behaves much more like a
mass dimension -$2$ quantity than a dimensionless observable.

\subsection{Modified lattice method for $\bf{a_\mu^{(2)}}$}

Having diagnosed the difficulty in $a_\mu^{(2)}$, first with
dimensional analysis and then a model argument, and having provided a
clean definition of the problem using $d_\mathrm{eff}$, we can now
attempt to remedy it.  In the end, we will define a modified quantity
$a_{\overline{\mu}}^{(2)}$ that has the same physical limit as
$a_\mu^{(2)}$ yet satisfies
$d_\mathrm{eff}[a_{\overline{\mu}}^{(2)}]=0$.  Since the physical
values of both observables are the same, we can safely use either
quantity to perform the computation.  Furthermore, our observation
that $d_\mathrm{eff}=0$ will provide a theoretical explanation for why
$a_{\overline{\mu}}^{(2)}$ should lend itself to an easier
calculation.

Starting with the observation that
$d_\mathrm{eff}[a_\mu^{(2)}]\not=0$, we sought a minimal way to modify
$a_\mu^{(2)}$ to eliminate this unexpected dependence on the lattice
spacing.  This is caused by the fact that $m_\mu$ is an external scale
and not capable of absorbing the dependence on $a$.  The solution we
came upon was to insert the factor $H_\mathrm{phys}^2 / H^2$ as
follows
\be
\label{albar}
a_{\overline{\mu}}^{(2)}
= \alpha^2 \int_0^\infty \frac{dQ^2}{Q^2} \omega\left( \frac{Q^2}{H^2} \cdot \frac{H_\mathrm{phys}^2}{m_\mu^2}\right) \Pi_R(Q^2)
=\alpha^2 \int_0^\infty \frac{d\hat{Q}^2}{\hat{Q}^2} \omega\left( \frac{\hat{Q}^2}{\hat{H}^2} \cdot \frac{H_\mathrm{phys}^2}{m_\mu^2}\right) \Pi_\mathrm{lat}(\hat{Q}^2)\,,
\ee
where $H$ is some hadronic scale and $H_\mathrm{phys}$ is its physical
limit value in physical units.  Additionally, the value of $H$ is
understood as being calculated at the same $m_{PS}$ as
$a_{\overline{\mu}}^{(2)}$.  Each choice of $H$ gives rise to a
distinct new observable $a_{\overline{\mu}}^{(2)}$.  We will shortly
pick a favored $H$, so the dependence on $H$ in defining
$a_{\overline{\mu}}^{(2)}$ is suppressed.  By construction, this
quantity has the same physical limit as the standard definition and
eliminates the unwanted dependence on the lattice spacing,
\bd
\lim_{m_{PS}\rightarrow m_\pi} a_{\overline{\mu}}^{(2)} = a_{\mu}^{(2)}
~~~~~~~~\mathrm{and}~~~~~~~~
d_\mathrm{eff}[a_{\overline\mu}^{(2)}]=0\,.
\ed
This modification completely eliminates the unwanted $a$ dependence,
but it does not automatically weaken the quark mass dependence.  We
know that the vector-mesons make a dominant contribution to
$a_\mu^{(2)}$ of the form $g_V^2 m_\mu^2 / m_V^2$ and that $g_V$,
shown in figure~\ref{gv}, is only weakly $m_{PS}$ dependent.  This
suggests choosing $H=m_V$.  Other choices have been examined
in~\cite{Feng:2011zk} but we will use $H=m_V$ exclusively in these
proceedings when discussing the modified approach.

The results for $a_{\overline\mu}^{(2)}$ are given in
figure~\ref{amubar}.  As $a_{\overline\mu}^{(2)}$ is now a proper
dimensionless observable composed of only QCD scales, it behaves like
any other dimensionless quantity, as we can check by comparing with
the $g_V$ in figure~\ref{gv}.  Additionally, the choice $H=m_V$
absorbs much of the $m_{PS}$ dependence that has troubled previous
calculations using the standard method.  All indications are that the
systematic errors are relatively mild.  Additionally, we have checked
that disconnected diagrams do not rise above the statistical errors
shown in figure~\ref{amubar}.  The extrapolated value is given in
table~\ref{nf-dep} and is consistent with the estimated $n_f=2$ piece
of the experimental determination.  More importantly, it is
encouraging that the resulting error on the physical limit value of
$a_\mu^{(2)}$ is already within a factor of $3\!-\!4$ of the
experimental determination using equation~\ref{amunf}.

\subsection{Comparison of current lattice calculations of $\bf{a_\mu^{(2)}}$}

\begin{figure}
\begin{minipage}{191pt}
\includegraphics[width=191pt]{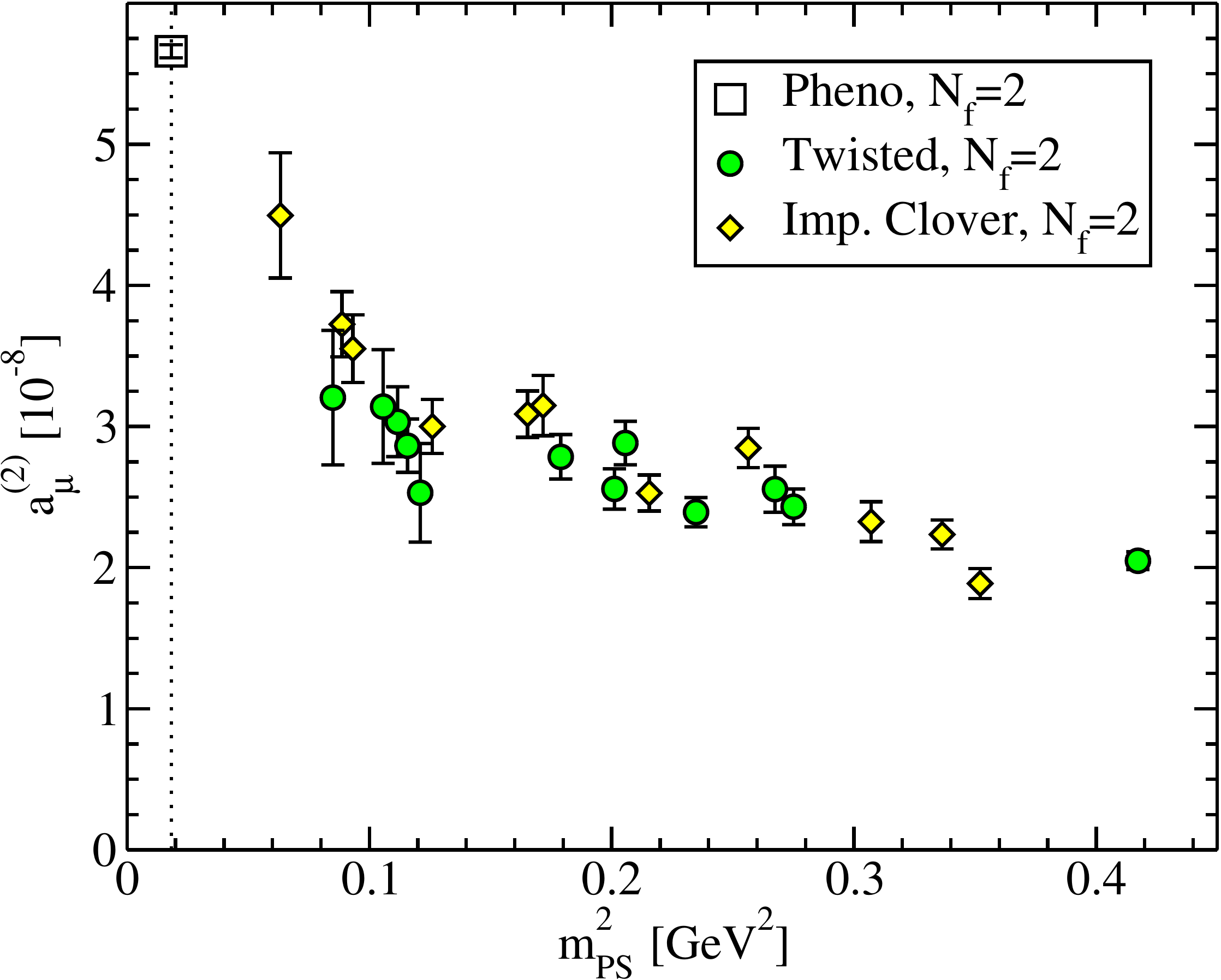}
\caption{Two-flavor lattice calculations of $a_\mu^{(2)}$.  The
  calculations are from \cite{Feng:2011zk} ("Twisted") and
  \cite{DellaMorte:2011aa} ("Imp. Clover").}
\label{nf2-comp}
\end{minipage}
\hspace{44pt}
\begin{minipage}{191pt}
\includegraphics[width=191pt]{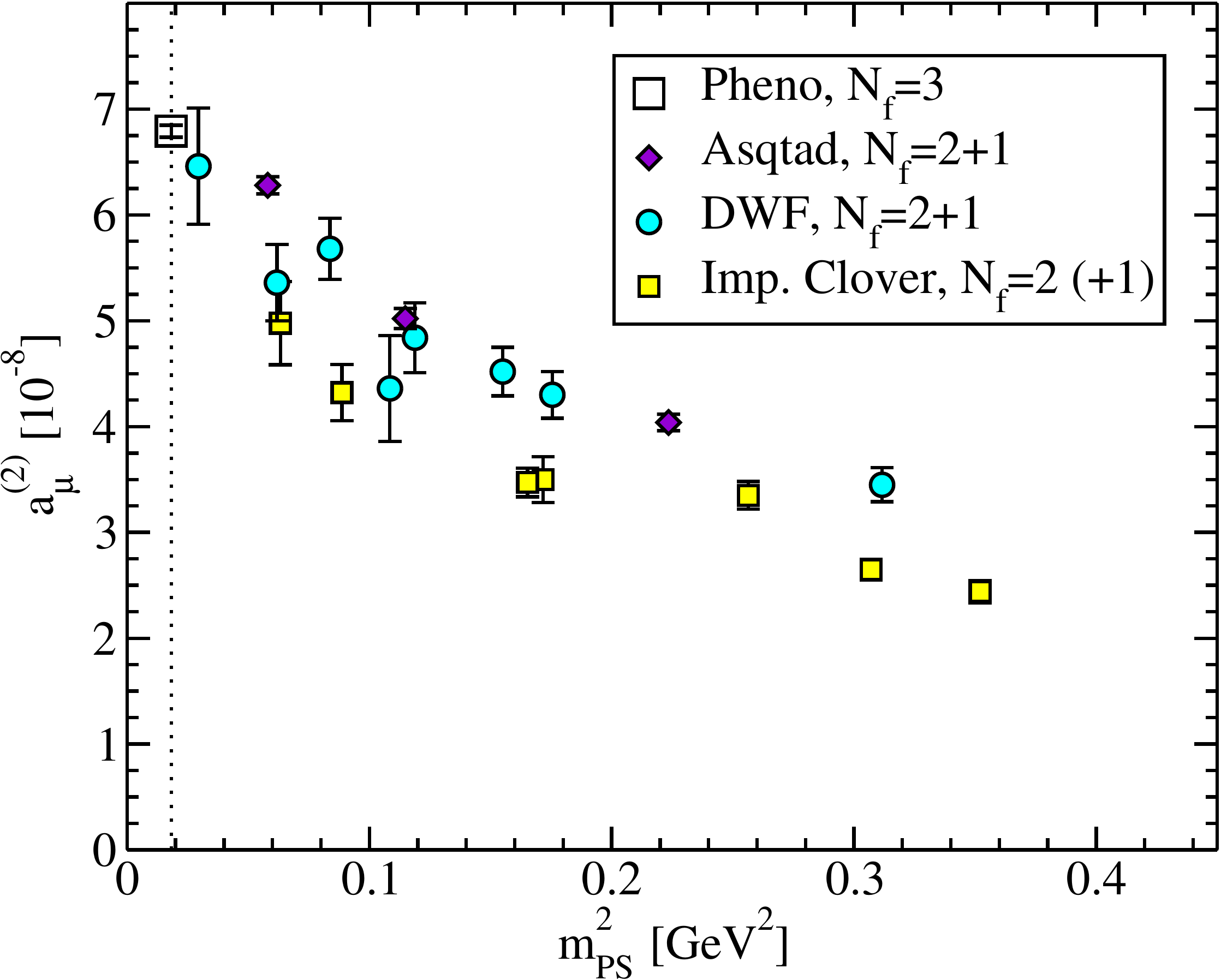}
\caption{Three-flavor lattice calculations of $a_\mu^{(2)}$.  The
  results are from \cite{Aubin:2006xv} ("Asqtad"), \cite{Boyle:2011hu}
  ("DWF") and \cite{DellaMorte:2011aa} ("Imp. Clover").}
\label{nf3-comp}
\end{minipage}
\end{figure}

Several groups have performed calculations of $a_\mu^{(2)}$ with both
$n_f=2$~\cite{Feng:2011zk,DellaMorte:2011aa} and
$n_f=3$~\cite{Aubin:2006xv,Boyle:2011hu}, and a first $n_f=4$
calculation is underway~\cite{grit}.  To compare results, we focus on
the standard method for this section.  First, we can examine the
lattice calculations for $n_f=2$, which are shown in
figure~\ref{nf2-comp}.
There we find agreement between both computations, including their
extrapolated values, which are given in table~\ref{nf-dep}.  These
results were calculated using different actions, at least two lattice
spacings, multiple physical volumes, a broad range of pion masses, and
different treatments of the low $Q^2$ extrapolations of $\Pi(Q^2)$.
The level of agreement is rather compelling for these two
calculations.  Additionally, it appears that the calculation of
\cite{DellaMorte:2011aa} shows a rapid rise as $m_{PS}$ is lowered,
which is consistent with the model-dependent expectations from the
vector-meson contribution discussed previously.

We now turn our attention to the $n_f=3$ calculations shown in
figure~\ref{nf3-comp}.  The situation here appears to be less
compelling than for $n_f=2$, but we must take some care before
reaching such a conclusion.  A detailed comparison requires a bit more
space than allowed in these proceedings, so we will limit ourselves to
commenting on one conceptual issue that is important to understand
before making any definitive statements.  When comparing the $n_f=3$
results, we must consider how the strange quark mass is determined.
Unless the chosen renormalization conditions match, there is no reason
why these curves would generally agree.  The only expectation is that
the values extrapolated to $m_{PS}=m_\pi$ must agree when all other
uncertainties have also been controlled for.  This seems to be the
case in figure~\ref{nf3-comp}.  Additionally, the extrapolated values
given in \cite{Boyle:2011hu} and \cite{DellaMorte:2011aa} (also given
in table~\ref{nf-dep}) do agree.  (The work of \cite{Aubin:2006xv} did
not cite a final result but it overlaps with the calculation of
\cite{Boyle:2011hu} for all three values of $m_{PS}$ used in
\cite{Aubin:2006xv}.)  Further understanding is needed for $n_f=3$,
but the results of the lattice calculations may be more encouraging
than is reflected in a simple head-to-head comparison as done in
figure~\ref{nf3-comp}.

\section{Leading-order QCD correction to the electron magnetic moment}

We now turn our discussion to a sequence of examples illustrating the
application of the modified method of~\cite{Feng:2011zk}.  The first
two examples are simple extensions to the other two leptons but
nonetheless provide nontrivial tests of the new method.  Besides,
these are first lattice calculations of both quantities.
In~\cite{Feng:2011zk}, the leading-order QCD contributions to the
electron and tau magnetic moments were also calculated, $a_e^{(2)}$
and $a_\tau^{(2)}$.  The electron magnetic moment is measured to a
precision of $0.28$ parts-per-trillion~\cite{Hanneke:2008tm}.  The
lightness of the electron makes $a_e$ significantly less dependent on
QCD corrections but the enhanced precision of the experimental
measurements overcomes the reduced sensitivity.  The current
measurement of $a_e$ is so precise that it is in fact used to
determine $\alpha$.  The standard model is then tested by comparing
this value to other determinations of $\alpha$.  These comparisons
have reached the precision that QCD effects of $a_e^{(2)}$ can not be
ignored.  However, the precision on $a_e$ is not yet high enough to
probe the error on $a_e^{(2)}$, so there is no pressing experimental
need for higher precision $a_e^{(2)}$ determinations.

\begin{figure}
\begin{minipage}{191pt}
\includegraphics[width=191pt]{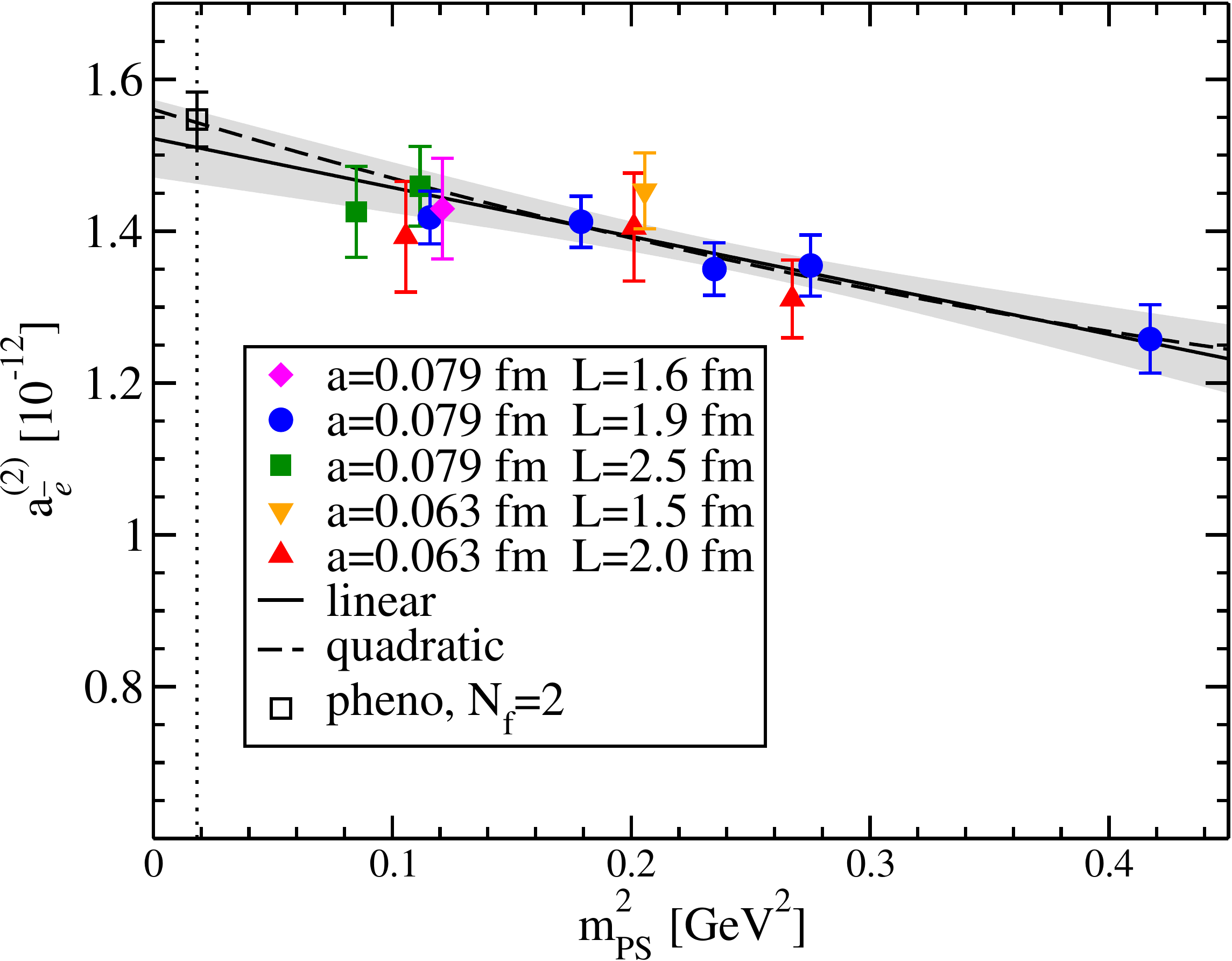}
\caption{Modified method $a_{\overline{e}}^{(2)}$.}
\label{aebar}
\end{minipage}
\hspace{44pt}
\begin{minipage}{191pt}
\includegraphics[width=191pt]{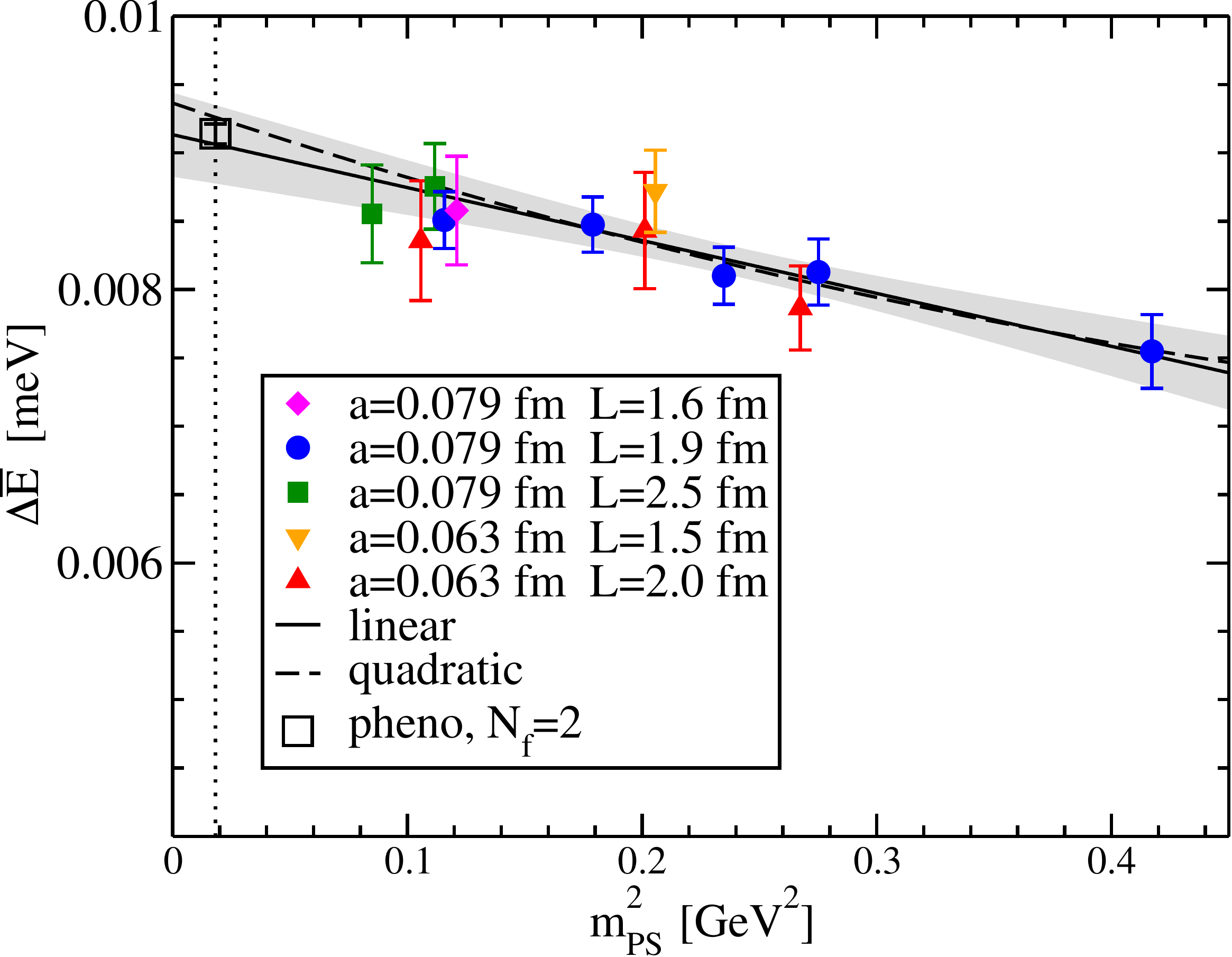}
\caption{Modified method $\Delta\overline{E}$.}
\label{mup}
\end{minipage}
\end{figure}
For our purposes, $a_e^{(2)}$ provides a test of the modified method
that is sensitive to only the extreme lowest $Q^2$ scales.  To a high
accuracy, $a_e^{(2)}$ can be approximated as
\bd
a_e^{(2)} = \frac{4}{3} \alpha^2 m_e^2 \left. \frac{d \Pi_R(Q^2) }{d Q^2} \right|_{Q^2=0} + {\cal O}(m_e/\Lambda)^4\,.
\ed
This approximation is not used in~\cite{Feng:2011zk}, but we mention
it in order to illustrate that $a_e^{(2)}$ probes essentially just the
derivative of $\Pi(Q^2)$ at $Q^2=0$.  It also illustrates in a simple
way again the idea that $a_e^{(2)}$ may behave differently from a
typical dimensionless observable.  In the case of $a_e^{(2)}$, the
approximation above strongly suggests that it will behave very much
like a mass dimension -$2$ observable.  This can be made precise by
noting that
\bd
d_\mathrm{eff}[a_e^{(2)}] = -1.999984\,(1)\,.
\ed
The results for $a_e^{(2)}$ from~\cite{Feng:2011zk} are shown in
figure~\ref{aebar} using the modified method.  The behavior of
$a_e^{(2)}$ is very similar to $a_\mu^{(2)}$ and agrees with the
estimated $n_f=2$ piece of the experimental measurement.

Additionally, the leading QCD vacuum-polarization correction to the
$2P\!\!-\!\!2S$ Lamb shift of muonic-hydrogen is also proportional to
the slope of $\Pi(Q^2)$ at $Q^2=0$ and hence closely related to
$a_e^{(2)}$.  This correction is given in~\cite{Faustov:1999fg}, with
$m_r$ the reduced mass of the $\mu\!\!-\!\!p$ system, as
\begin{displaymath}
\Delta E = 2 \pi \alpha^5 m_r^3 \left. \frac{d\Pi_R}{dQ^2} \right|_{Q^2=0}
~~~~\mathrm{and}~~~~
\Delta\overline{E} = 2 \pi \alpha^5 m_r^3 \left. \frac{d\Pi_R(Q^2/H_\mathrm{phys}^2\cdot H^2)}{dQ^2} \right|_{Q^2=0}
\,.
\end{displaymath}
The second form is the modified approach for $\Delta E$ and results
from a consistent treatment of external scales as discussed thoroughly
in section~\ref{sec:dalpha}.  The results for $\Delta\overline{E}$ are
shown in figure~\ref{mup}.

\section{Leading-order QCD correction to the tau magnetic moment}

\begin{figure}
\begin{minipage}{191pt}
\includegraphics[width=191pt]{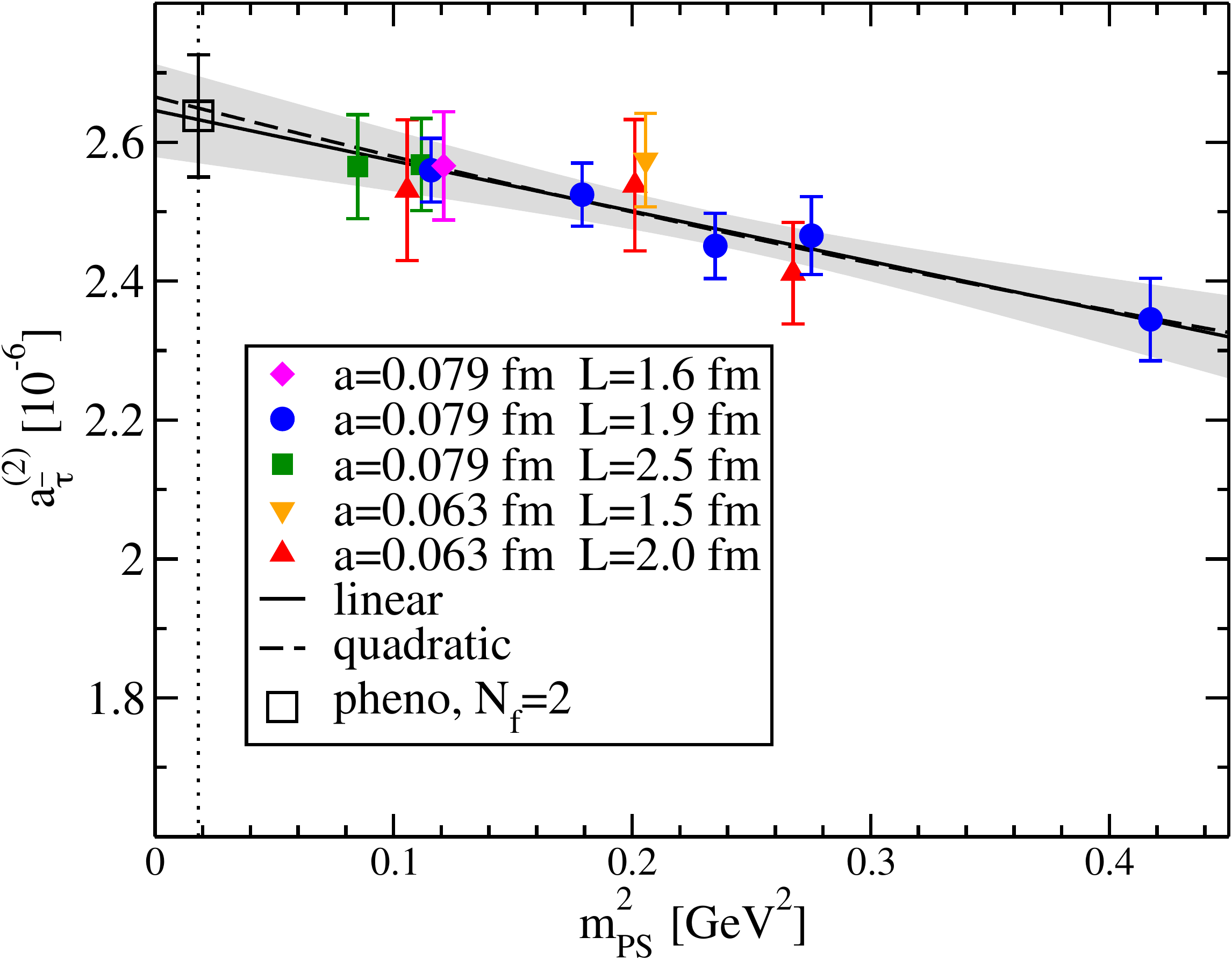}
\caption{Modified method $a_{\overline{\tau}}^{(2)}$.}
\label{ataubar}
\end{minipage}
\hspace{44pt}
\begin{minipage}{191pt}
\includegraphics[width=191pt]{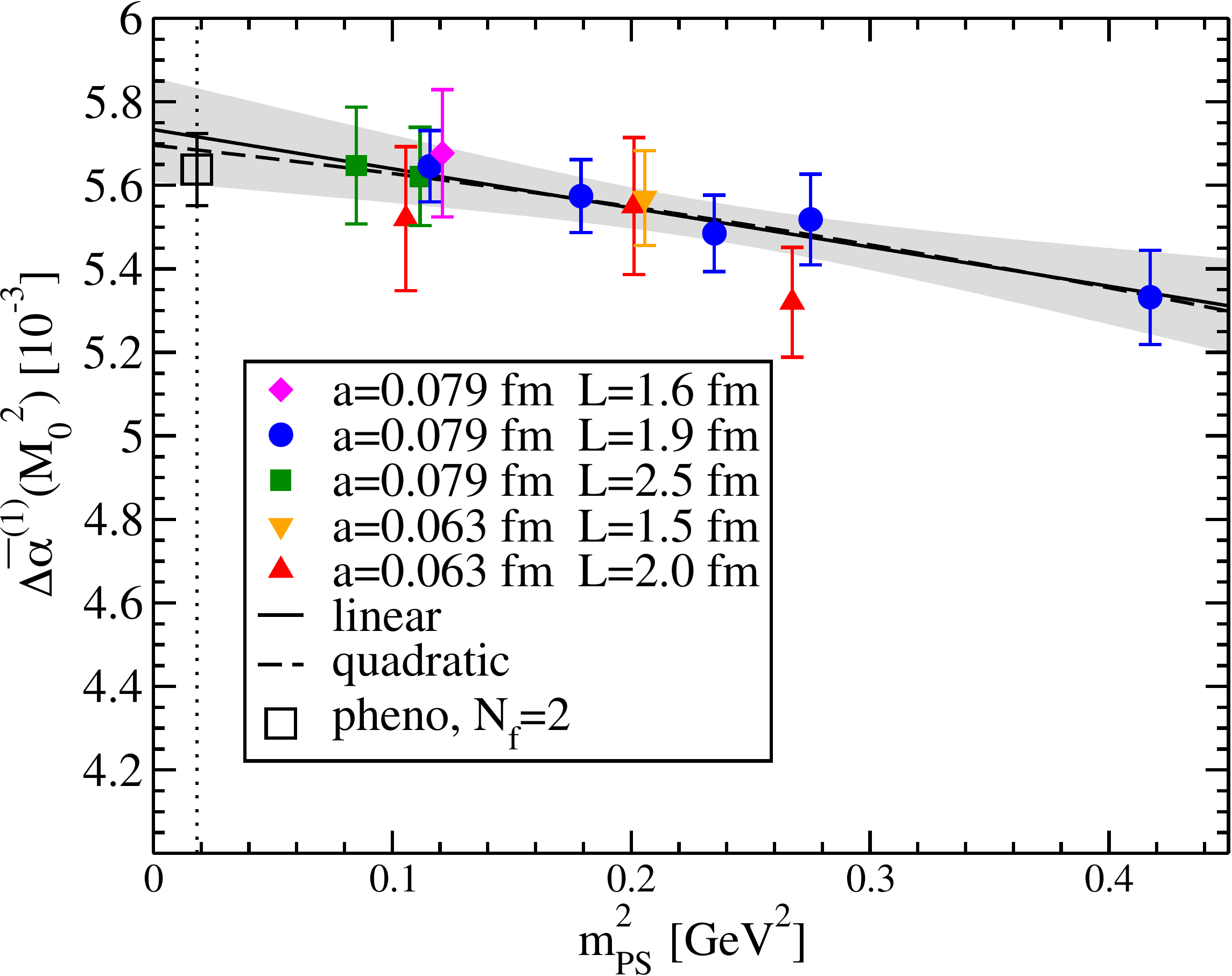}
\caption{Modified method $\Delta\overline{\alpha}^{(1)}(M_0^2)$.}
\label{dalpha_s0}
\end{minipage}
\end{figure}

The magnetic moment of the tau is substantially more sensitive to
potential new physics than that of the muon, but it has not been
experimentally measured.  There are experimental bounds on
$a_\tau$~\cite{Nakamura:2010zzi}, but more interesting to us is that
$a_\tau^{(2)}$ can be determined by the same analysis used for
$a_\mu^{(2)}$ and $a_e^{(2)}$.  Additionally, due to the heaviness of
the tau, it is sensitive to a different range of QCD scales.  In fact,
\bd
d_\mathrm{eff}[a_\tau^{(2)}] = -0.936\,(13)
\ed
indicates that it behaves quite a bit different than $a_\mu^{(2)}$ and
$a_e^{(2)}$ and more like a mass dimension -$1$ observable.  In fact,
applying the standard method to $a_\tau^{(2)}$ leads to a reasonable
calculation.  However, the arguments for the modified method still
apply here and it is yet another test of the approach.  The results
for $a_{\overline\tau}^{(2)}$ are shown in figure~\ref{ataubar}.
Again, the resulting observable has a mild $m_{PS}$ dependence.
Furthermore, it agrees with the estimated $n_f=2$ piece of the
dispersive result, determined from equation~\ref{amunf}, providing
more confidence in both the modified method and the simple
prescription for analyzing the $n_f$ dependence of $R(s)$.  As
figure~\ref{ataubar} shows, the lattice determination of
$a_\tau^{(2)}$ is already more accurate than its experimental
determination, suggesting that future measurements of $a_\tau$ can in
fact rely on a fully nonperturbative determination of $a_\tau^{(2)}$
without sacrificing any precision once $n_f=4$ lattice calculations
have been completed.

\section{Leading-order QCD contribution to the running of the QED coupling}
\label{sec:dalpha}

In order to further demonstrate and understand the modified method
of~\cite{Feng:2011zk}, we have applied the same idea to the
determination of the leading-order QCD corrections to the running of
$\alpha$.  This is the first lattice calculation of this quantity and
all results discussed in these proceedings are preliminary.  The
running of $\alpha$ is normally treated by introducing an effective
coupling given by summing all one-particle irreducible bubble
insertions in the photon propagator.  This results in
\bd
\alpha(Q^2) = \frac{\alpha}{1-\Delta\alpha(Q^2)}\,.
\ed
The QCD contribution $\Delta\alpha^{\mathrm{QCD}}$ is again expanded
in $\alpha$.  The leading-order correction is
\bd
\Delta\alpha^{(1)}(Q^2) = 4 \pi \alpha\, \Pi_R(Q^2)\,.
\ed
The value of $\alpha(Q^2=0)$ is just the usual coupling $\alpha$,
which is known to a relative precision of $1\cdot 10^{-9}$.  However,
after evolving $\alpha$ to a high scale, say the $Z$-boson pole at
$Q^2=M_Z^2$, the relative precision on $\alpha(M_Z^2)$ drops to
$1\cdot 10^{-4}$~\cite{Davier:2010nc}, making $\alpha(M_Z^2)$ one of
the more poorly known fundamental parameters in high energy
predictions.  Similar to $a_\mu$, the dominant uncertainty in this
evolution is due to hadronic corrections, which is then passed into
every high energy process through the use of the running coupling
$\alpha(M_Z^2)$.  This has larger impact than one might have naively
expected.  As one example, a recent global analysis by the Gfitter
collaboration determined the Higgs mass to be
$m_H=44^{+62}_{-43}~\mathrm{GeV}$, if the experimental determination
of $\Delta\alpha(M_Z^2)$ was not included in the fit and found
$m_H=96^{+31}_{-24}~\mathrm{GeV}$ if it was~\cite{Baak:2011ze}.

The treatment of the external scales for $a_l^{(2)}$ introduced
earlier uniquely fixes the treatment of the now external scale $Q^2$
in $\Delta\alpha(Q^2)$.  To see this, we can simply rewrite
equation~\ref{albar} as
\bd
a_{\overline{l}}^{(2)} =
\alpha^2 \int_0^\infty \frac{dQ^2}{Q^2} \omega\left( Q^2 / m_l^2 \right) \Pi_R\left( Q^2 / H_\mathrm{phys}^2 \cdot H^2\right)\,.
\ed
This suggests rather clearly that we should consider the following
modified definition of $\Delta\alpha^{(1)}$
\bd
\Delta\overline{\alpha}^{(1)}(Q^2) \equiv
4 \pi \alpha\, \Pi_R(Q^2 / H_\mathrm{phys}^2 \cdot H^2 )\,.
\ed
Just as for $a_{\overline{l}}^{(2)}$, this new quantity explicitly has
the correct physical limit but also satisfies $d_\mathrm{eff} = 0$.

We can examine the consequences of this definition by first focusing
on $Q^2=M_0^2$ with $M_0=2.5~\mathrm{GeV}$, which is a common matching
scale for phenomenological work.  The lattice calculation is shown in
figure~\ref{dalpha_s0}.  We see that, just as for each of the
$a_l^{(2)}$ calculations, the modified definition results in a rather
mild looking extrapolation to the physical point, giving
$\Delta\alpha^{(1)}(M_0^2)=5.72\,(12)\cdot 10^{-3}$.  We can also
apply the same treatment of the $n_f$ dependence of the experimentally
determined $R(s)$ to $\Delta\alpha^{(1)}$ resulting in
$\Delta\alpha^{(1)}(M_0^2)=5.60\,(06)\cdot 10^{-3}$.  The preliminary
lattice computation results in an uncertainty that is now only twice
the error of the experimentally determined quantity, suggesting that
lattice calculations could be a competitive, if not superior, way to
determine $\Delta\alpha(Q^2)$ at least for low scales.

\begin{figure}
\begin{minipage}{191pt}
\includegraphics[width=191pt]{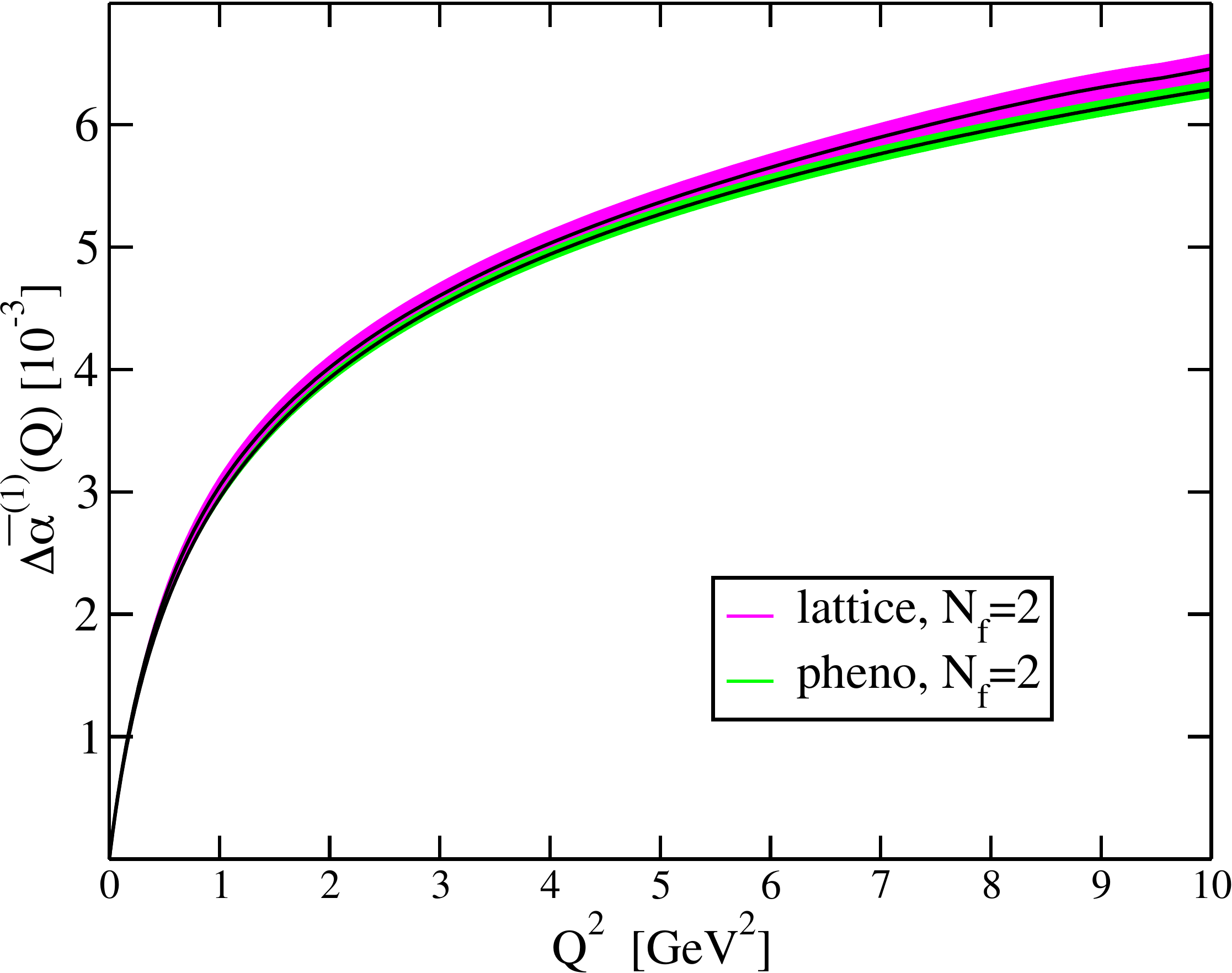}
\caption{Modified method $\Delta\overline{\alpha}^{(1)}(Q^2)$.}
\label{dalpha}
\end{minipage}
\hspace{44pt}
\begin{minipage}{191pt}
\includegraphics[width=191pt]{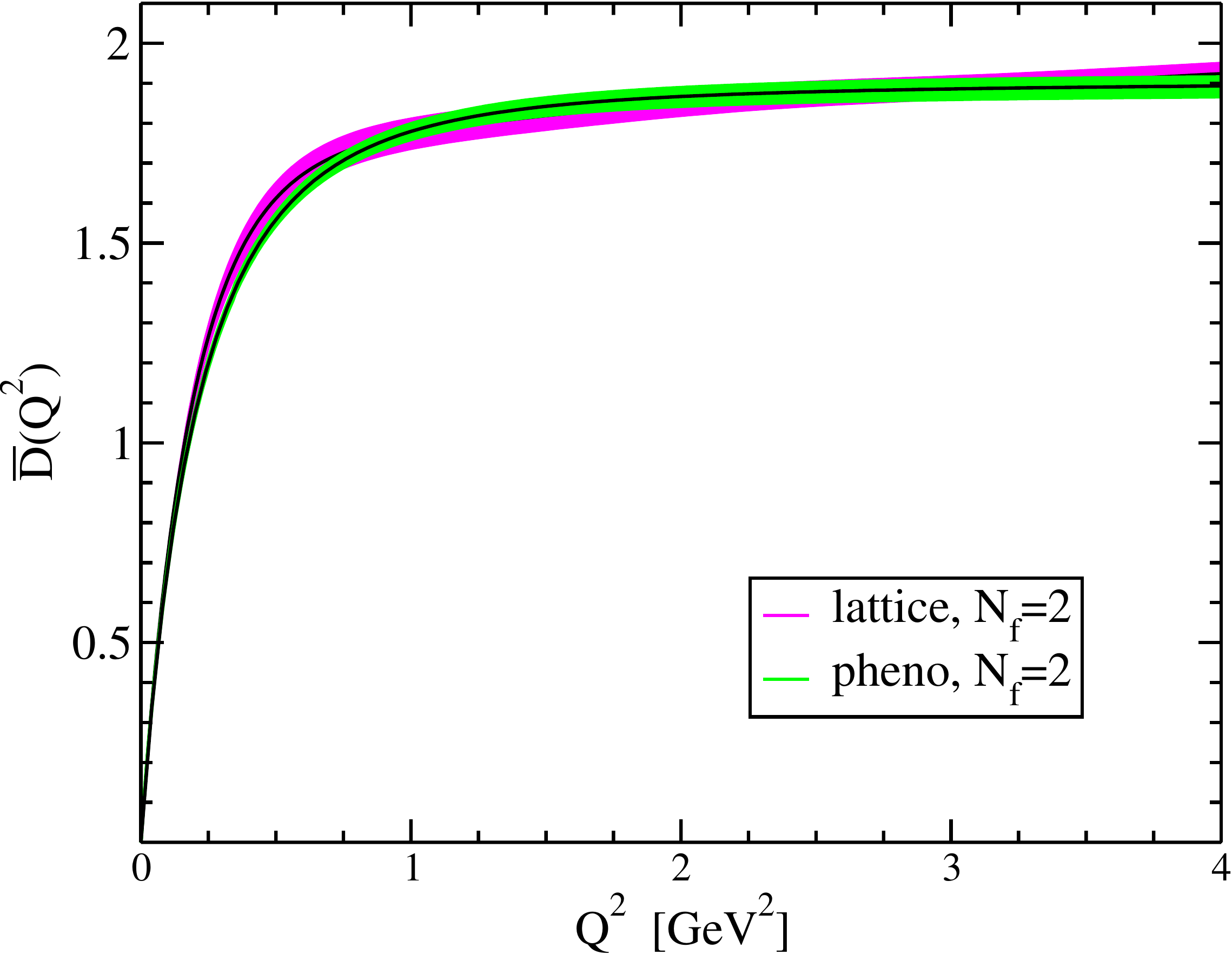}
\caption{Modified method $\overline{D}(Q^2)$.}
\label{adler}
\end{minipage}
\end{figure}

Now, we can repeat the analysis for all $Q^2$ and determine the QCD
induced running of $\alpha(Q^2)$.  This is shown in
figure~\ref{dalpha}.
By comparing results with different lattice spacings, we notice
significant lattice artifacts only for $Q^2\gtrsim 7~\mathrm{GeV}^2$.
This appears to be just a mild obstacle to an accurate determination
of $\Delta\alpha(Q^2)$ in the relevant low $Q^2$ regime.  The scale of
$Q^2=M_0^2$ was chosen because perturbation theory becomes reliable
for yet larger $Q^2$.  To run $\alpha$ to higher scales, we have
determined $\alpha_s$ by matching $\Pi(Q^2)$ to the perturbative
expectations for the $Q^2$ regions that can be reached in lattice
calculations and then determined $\Delta\alpha^{(1)}(Q^2)$ at larger
scales through
\bd
\Delta\alpha^{(1)}(M_Z^2) =
\Delta\alpha^{(1)}(M_0^2) + ( \Delta\alpha^{(1)}(M_Z^2) - \Delta\alpha^{(1)}(M_0^2) )\,,
\ed
where the perturbative expression for $\Delta\alpha^{(1)}(M_Z^2) -
\Delta\alpha^{(1)}(M_0^2)$ is available at $5$
loops~\cite{Baikov:2008jh}.

We note that to determine $\alpha_s$, it appears to be better to
calculate the Adler function $D(Q^2)=d \Pi(Q^2) / d \ln(Q^2)$.  To
consistently apply our treatment of the external scale $Q$, we define
a modified Adler function
\begin{displaymath}
\overline{D}(Q^2) = D(Q^2/H_\mathrm{phys}^2 \cdot H^2)\,.
\end{displaymath}
The results of a preliminary calculation with the modified technique
are shown in figure~\ref{adler}.

\section{Next-to-leading-order QCD contribution to the muon magnetic moment}

The precision of the current BNL measurement of $a_\mu$ already
requires that the next-to-leading-order QCD correction $a_\mu^{(3)}$
be accounted for.  Most of the diagrams involve insertions of the
vacuum-polarization correction into lower order QED diagrams, but a
new QCD contribution, called light-by-light, also occurs at this
order.

\subsection{Vacuum-polarization corrections}

\begin{figure}
\begin{minipage}{191pt}\hspace{15pt}
\includegraphics[width=100pt]{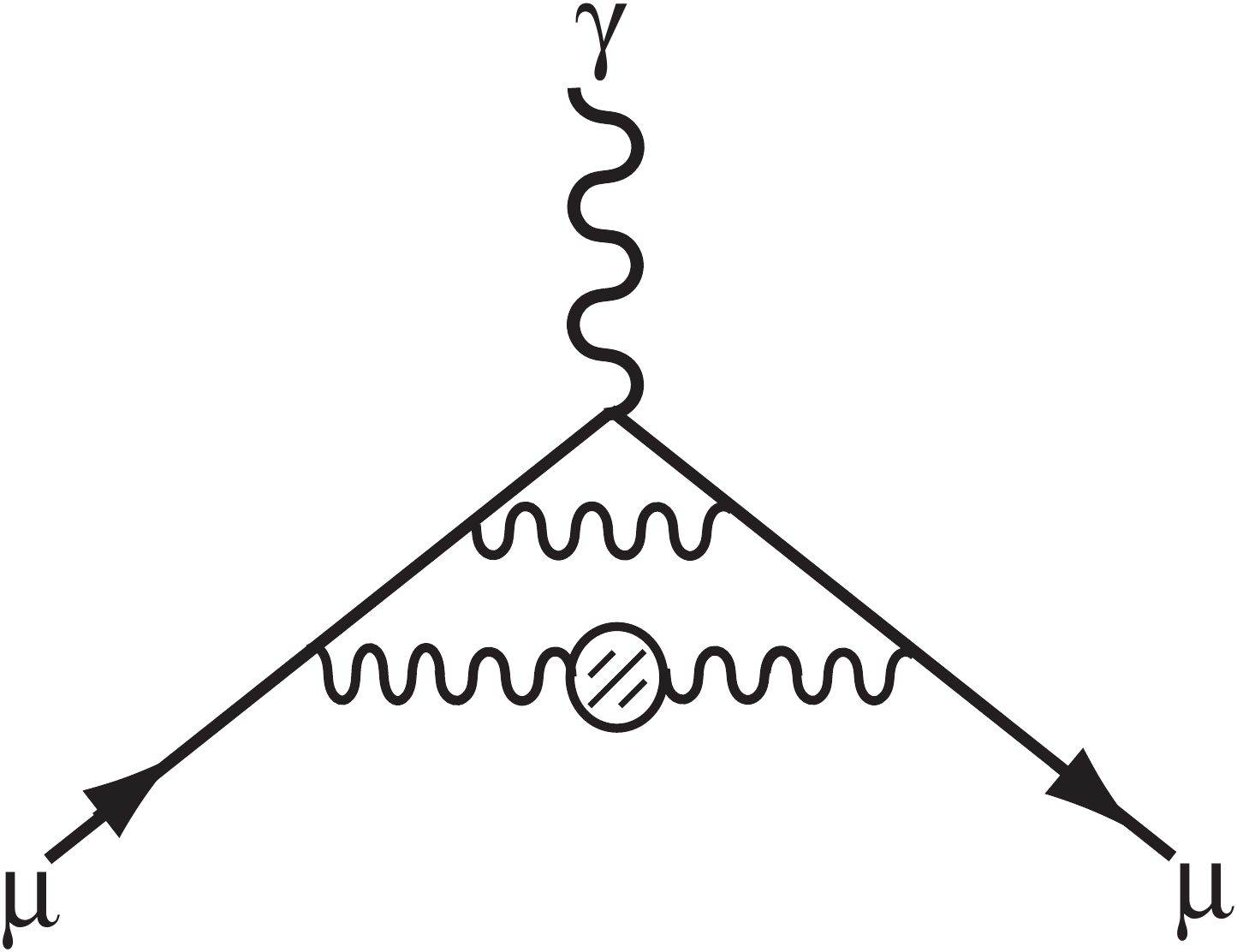}

\hspace{91pt}\includegraphics[width=100pt]{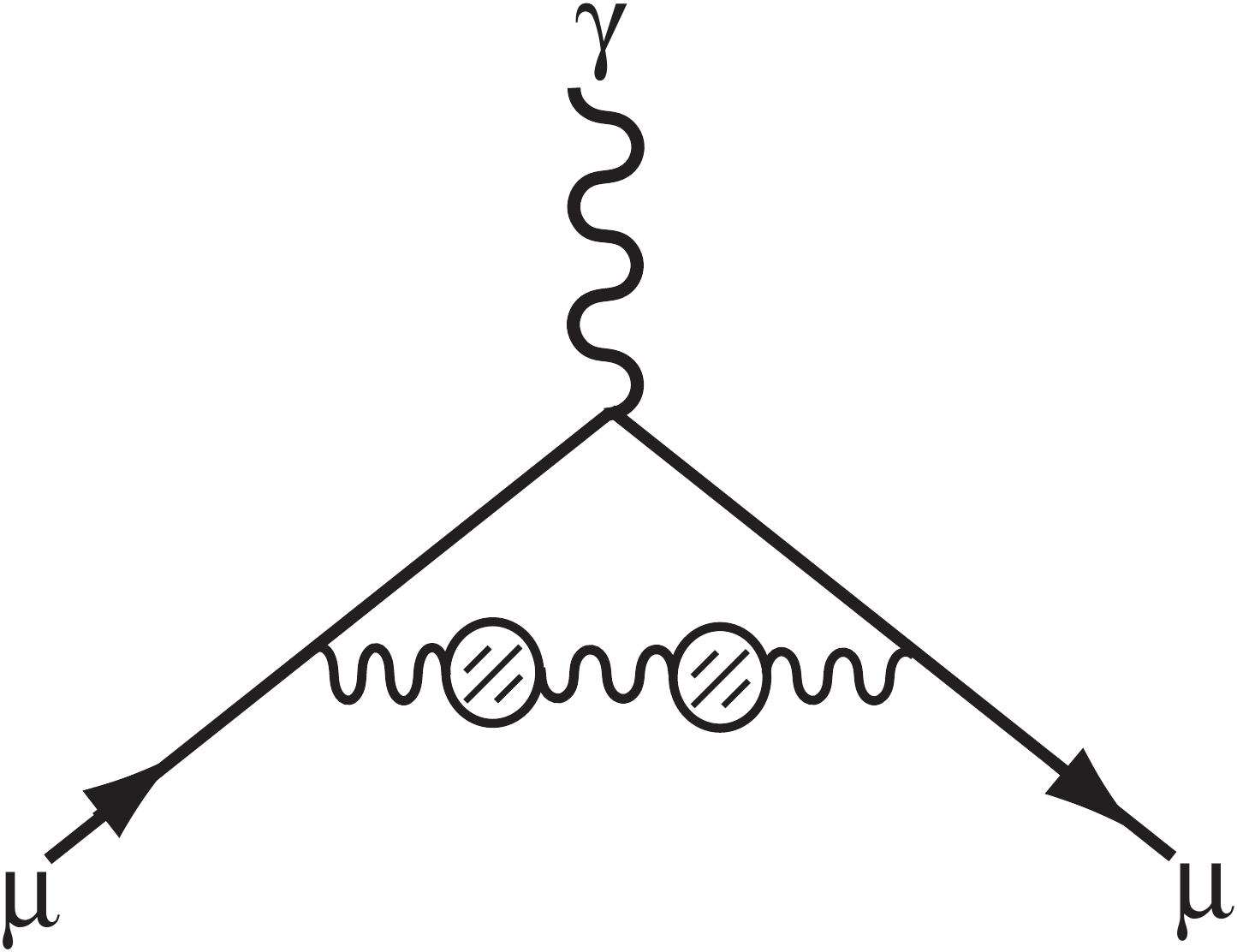}
\caption{Example diagrams for $a_\mu^{(3,\mathrm{vp})}$.}
\label{hovp}
\end{minipage}
\hspace{44pt}
\begin{minipage}{191pt}
\includegraphics[width=191pt]{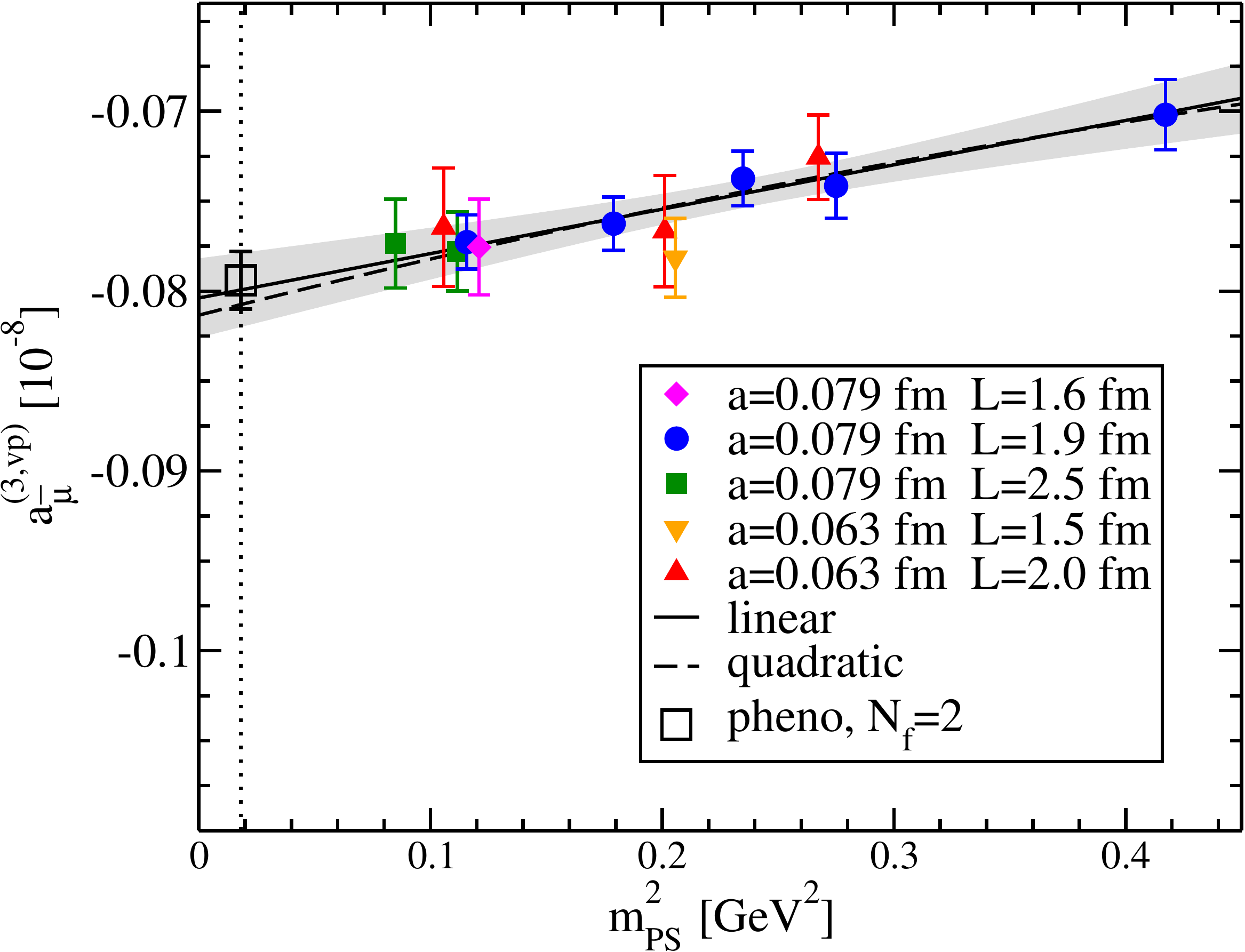}
\caption{Modified method $a_{\overline\mu}^{(3,\mathrm{vp})}$.}
\label{amubarho}
\end{minipage}
\end{figure}

The vacuum-polarization correction can be inserted once or twice into
any photon line of a two-loop or one-loop QED diagram, respectively.
This results in 16 diagrams that involve one occurrence of $\Pi(Q^2)$
and one diagram with two insertions of $\Pi(Q^2)$.  Example diagrams
are shown in figure~\ref{hovp}.
Expressions in various forms are available for these corrections when
expressed as integrals over $R$.  The analytic continuation to
Euclidean space is complicated for some of these contributions, but it
appears that all vacuum-polarization contributions to $a_\mu^{(3)}$
can be calculated in Euclidean space.  A preliminary calculation using
the modified approach\footnote{At the conference, only a partial
  accounting of the diagrams was given and resulted in a different
  value.  The result reported here accounts for all
  vacuum-polarization contributions to $a_\mu^{(3)}$.}, shown in
figure~\ref{amubarho}, gives
$a_\mu^{(3,\mathrm{vp})}=-7.99\,(20)\cdot10^{-10}$.  The uncertainty
includes only statistical errors but the systematic uncertainties
appear to be small.  For comparison, the $n_f=2$ piece of the
experimental measurement is
$a_\mu^{(3,\mathrm{vp})}=-7.94\,(16)\cdot10^{-10}$.  Further study is
underway, but the initial results for the vacuum-polarization
contribution to $a_\mu^{(3)}$ seem to agree with the expectations from
the experimental measurements and have a nearly comparable
uncertainty.  Most importantly, the precision of the lattice result is
better than the accuracy expected for the future muon $g\,$-$\,2$
measurements, so it seems that lattice QCD should be quite capable of
determining the higher-order vacuum-polarization contributions at the
required precision.

\subsection{Light-by-light corrections}

In contrast to the vacuum-polarization corrections, the light-by-light
contributions to $a_\mu^{(3)}$ represent a real challenge.  There are
ongoing lattice studies by several
groups~\cite{Hayakawa:2005eq,Rakow,Shintani:2009qp}, each using
different methods but still all exploratory.  The uncertainty on the
light-by-light correction $a_\mu^{(3,\mathrm{lbl})}$ is nearly as
large as that of $a_\mu^{(2)}$ and there are open questions regarding
the methods currently used for its determination, so a nonperturbative
calculation of $a_\mu^{(3,\mathrm{lbl})}$ is highly desirable.
However, a lattice calculation of one piece of the higher-order
contribution is less satisfying than a complete lattice calculation of
both the leading-order and next-to-leading-order corrections.  It now
appears that the vacuum-polarization pieces should be calculable, so
the light-by-light contribution is the only remaining piece needed for
a completely nonperturbative determination of $a_\mu^\mathrm{QCD}$
accurate to ${\cal O}(\alpha^3)$.  We can only hope that this will
encourage an even greater effort within the lattice community to
tackle the desperately needed light-by-light contribution.

\section{Conclusions}

We have discussed several examples of important measurements that
receive sizable hadronic corrections.  The muon $g\,$-$\,2$, which
hints at beyond-the-standard-model physics, is the most pressing
observable, but $\Delta\alpha(Q^2)$ may in fact have a much broader
impact on precision standard model predictions.  Using a modified
lattice approach, the leading QCD corrections to both of these
observables appear to be reliably calculable.  To further explore the
new method, we have also examined the leading corrections to the
electron and tau leptons and the Lamb shift in muonic-hydrogen.  We
have calculated the Adler function, which can be matched to
perturbation theory to determine the strong coupling.  Lastly, we have
worked out methods to examine all the vacuum-polarization corrections
at the next-to-leading order.  In several cases, the currently reached
precisions on these quantities are approaching that of the
corresponding experimental determinations.  This indicates that fully
nonperturbative determinations of QCD corrections to electroweak
observables may be feasible at the precisions needed by future
experimental measurements that aim to discover or constrain physics
beyond the standard model.

\section{Acknowledgments}

We would like to thank the following colleagues for detailed
discussions regarding their lattice calculations of QCD corrections to
the muon anomalous magnetic moment:\ Christopher Aubin, Tom Blum,
Peter Boyle, Luigi Del Debbio, Bejamin J\"ager, Eoin Kerrane, Michele
Della Morte, Hartmut Wittig, and James Zanotti.  This manuscript has
been coauthored by Jefferson Science Associates, LLC under Contract
No.\ DE-AC05-06OR23177 with the U.S.\ Department of Energy.  X.~F.\ is
supported in part by the Grant-in-Aid of the Japanese Ministry of
Education (No.\ 21674002).  This work is supported in part by the DFG
Sonder\-for\-schungs\-be\-reich/Trans\-regio SFB/ TR9. HPC resources
were provided by the JSC Forschungszentrum J\"ulich on the JuGene
supercomputer.

\bibliographystyle{h-physrev}
\bibliography{lat11-renner.bib}

\end{document}